\documentclass[twocolumn,twoside]{revtex4}
\usepackage{graphicx}
\usepackage{fancyhdr}
\usepackage{float}

\def\beq{\begin{equation}}
\def\eeq{\end{equation}}
\def\bea{\begin{eqnarray}}
\def\eea{\end{eqnarray}}
\def\nn{\nonumber}
\def\roughly#1{\mathrel{\raise.3ex\hbox
{$#1$\kern-.75em\lower1ex\hbox{$\sim$}}}}

\def\bsmumu{b \to s \mu^+ \mu^-}

\def \hc{{\rm h.c.}}

\def \cB{{\cal B}}
\def \cL{{\cal L}}

\def \SM{{\rm SM}}

\def \De{\Delta}
\def \cL{{\cal L}}

\def \oQ{\overline{Q}}

\def \si{\sigma}
\def \ga{\gamma}
\def \({\left(}
\def \){\right)}
\def \[{\left[}
\def \]{\right]}
\def \<{\left<}
\def \>{\right>}
\def \l|{\left|}
\def \r|{\right|}

\begin{document}

\title{Combined explanation of the B-anomalies}

%

\author{Jacky Kumar}
\affiliation{Physique des Particules, Universit´e de Montr´eal,
C.P. 6128, succ. centre-ville, Montr´eal, QC, Canada H3C 3J7}

\begin{abstract}
There are four models of tree-level new physics (NP) that can potentially explain the $b\to s \mu^+ \mu^-$ and $b \to c\ell \bar \nu$ anomalies simultaneously. They are the S3, U3, and U1 leptoquarks and a standard-model-like triplet vector boson (VB). In this talk, I describe an analysis of these models with general couplings. We find that even in this most general case S3 and U3 are excluded. For the U1 model, I discuss the importance of the constraints from lepton- flavor-violating(LFV) processes. As for the VB model, it is shown to be excluded by the additional tree level constraints and LHC bounds on high-mass resonant dimuon pairs. This conclusion is reached without any assumptions about the NP couplings.
\end{abstract}

\maketitle

\thispagestyle{fancy}


Currently, there are a number of measurements of the B-decays which do not agree with the 
standard model(SM). The size discrepancies vary from $2$-$4 \sigma$ and the combined significance based on the global fits amounts roughly to $5$-$6\sigma$. Here is the list of individual anomalies:
\begin{itemize}
\item In $b$ $\to$ $ s \mu^+ \mu^-$ data, the measurements of the angular observables in $B\to K^* \mu^+ \mu^-$ and branching ratios in the decay $B_s \to \phi \mu^+\mu^-$ deviate from the SM at the level of $\sim  4\sigma$\cite{BK*mumuLHCb1, BK*mumuLHCb2, BK*mumuBelle, BK*mumuATLAS, BK*mumuCMS, BsphimumuLHCb1, BsphimumuLHCb2}.
\item  The letpton flavor universality (LFU) ratios $R_K$ and $R_{K^*}$ which are defined to  be ratio of the branching fractions for $\mu$ and $e$ modes in $B^+ \to K^+ \ell \ell$ and $B\to K^* \ell \ell$ decays respectively. Both are measured to be below the SM by about $2.5\sigma$\cite{RKexpt,RK*expt}.
\item   The LFU ratios in the charge current decay $B \to D^{(*)} \ell \bar \nu$, so called $R_D$ and $R_{D^*}$, deviate from SM at the level of $4\sigma$\cite{RD_BaBar, RD_Belle, RD_LHCb, Abdesselam:2016xqt}. Moreover, the $R_{J/\psi}$ also disagrees with the SM by about $2\sigma$\cite{Aaij:2017tyk}. 
\end{itemize}
In terms of the individual explanations of neutral and charged current anomalies, the best way is to use weak effective theory(WET), which is valid below the electroweak weak scale. The effective lagrangian containing dimension six terms reads
\begin{equation}
	\mathcal{L}_{eff}= -\mathcal{H}_{eff} = \sum_i C_i O_i
\end{equation}
here $C_i$, the wilson-coefficients(WCs) receive contributions from SM and NP. As far as $b\to s \ell \ell$ transitions are concerned a shift $C_9^{\mu\mu} =-C_{10}^{\mu\mu} \sim -0.53$ in the WC of operators $(\bar s \gamma^\mu P_L b)(\bar \mu \gamma_\mu \mu) $ and $ (\bar s \gamma^\mu P_L b) (\bar \mu \gamma_\mu \gamma^5 \mu)$ is sufficient\footnote{Although after the Moriond 2019 update of $R_K$ measurement\cite{Aaij:2019wad} this picture has slightly changed.}. 
For the charged current anomalies a NP contribution to the WC $C_V^{\tau \tau} \simeq 0.1$ of operator  $(\bar c \gamma^\mu P_L b)(\bar \tau \gamma_\mu P_L \nu)$ is needed \cite{Capdevila:2017bsm, Altmannshofer:2017yso, DAmico:2017mtc, Hiller:2017bzc, Geng:2017svp, Ciuchini:2017mik, Celis:2017doq, Alok:2017sui}. For the combined explanation Standard model effective theory(SMEFT) is best suitable. The SMEFT is built from the operators upto dimension six made of the SM fields respecting the full SM gauge symmetry\cite{Grzadkowski:2010es}. In SMEFT, the operator $(\bar Q_p \gamma_\mu \sigma^I P_L Q_r)(\bar L_s \gamma^\mu \sigma^I P_L L_t)$ relates the $b\to s$ to $b\to c$ transitions\cite{RKRD}. Restricting ourselves to the (V-A) structure, we also need to consider operator $(\bar Q_p \gamma_\mu P_L Q_r)(\bar L_s \gamma^\mu P_L L_t)$ which contribute to only to $b\to s\ell \ell$.

The next step is to look at the models which generate these two operators. Assuming that the SM is extended only by a single new particle, there are four possibilities. Three of them involve the Leptoquarks(LQs) and the fourth an addition vector boson. Three LQ models are scalar triplet(S3), vector triplet(U3) and vector singlet(U1). Here the transformation property refer to $SU(2)_L$ gauge group of SM.

Allowing the NP couplings to take general real values under the assumption that the NP couple only the second and third generations, a systematic analyses of these four models based on the paper\cite{Kumar:2018kmr} is presented here.

The Lagrangian of the LQ models read
\bea
\De\cL_{S_3} &=& h^{S_3}_{ij} \(\oQ_{iL}\si^I i\sigma^2 L^c_{j L}\)S^{I}_{3} + \hc, \nn\\
\De\cL_{U_3} &=& h^{U_3}_{ij} \(\oQ_{i L}~\ga^\mu~\si^I L_{j L}\)U^{I}_{3\mu} ~+~ \hc, \nn\\
\De\cL_{U_1} &=& h^{U_1}_{ij} \(\oQ_{i L}~\ga^\mu~L_{j L}\) U_{1\mu} + \hc
\eea
Each model can be described by four real couplings $h_{ij}$, here i and j can take values 2-3. The analysis is focused on the two things, first to find out which models work and second to analyze the pattern of the NP couplings for that purpose the LFV constrained turned out to be instrumental. The complete list of the observables playing an important role are shown in Table \ref{tab:obs_meas}. The observables are divided into two categories which we refer as \emph{minimal constraints} and \emph{LFV constraints}.  
\begin{table*}[t]
\begin{center}
\begin{tabular}{|c|c|}
 \hline
Observable & Measurement or Constraint \\
\hline
minimal & \\
\hline
$\bsmumu$ (all) & $C_9^{\mu\mu}({\rm LQ}) = -C_{10}^{\mu\mu}({\rm LQ}) = -0.68 \pm 0.12$ \cite{Alok:2017sui} \\
$R_{D^*}^{\tau/\ell}/(R_{D^*}^{\tau/\ell})_\SM$ & $1.18 \pm 0.06$ \cite{RD_BaBar, RD_Belle, RD_LHCb,Abdesselam:2016xqt} \\
$R_{D}^{\tau/\ell}/(R_{D}^{\tau/\ell})_\SM$ & $1.36 \pm 0.15$ \cite{RD_BaBar, RD_Belle, RD_LHCb,Abdesselam:2016xqt} \\
$R_{D^*}^{e/\mu}/(R_{D^*}^{e/\mu})_\SM$ & $1.04 \pm 0.05$ \cite{Abdesselam:2017kjf} \\
$R_{J/\psi}^{\tau/\mu}/(R_{J/\psi}^{\tau/\mu})_\SM$ & $2.51 \pm 0.97$ \cite{Aaij:2017tyk} \\
$\cB(B \to K^{(*)} \nu {\bar\nu})/\cB(B \to K^{(*)} \nu {\bar\nu})_\SM$ &
        $-13 \sum_{i=1}^3 {\rm Re}[C_L^{ii}({\rm LQ})] + \sum_{i,j=1}^3 |C_L^{ij}({\rm LQ})|^2 \le 248$ \cite{Alok:2017jgr} \\
\hline
LFV & \\
\hline
$\cB(B^+ \to K^+ \tau^- \mu^+)$ & $(0.8 \pm 1.7) \times 10^{-5}$ ~;~~ $< 4.5 \times 10^{-5}$ (90\% C.L.)  \cite{Lees:2012zz} \\
$\cB(B^+ \to K^+ \tau^+ \mu^-)$ & $(-0.4 \pm 1.2) \times 10^{-5}$ ~;~~ $< 2.8 \times 10^{-5}$ (90\% C.L.) \cite{Lees:2012zz} \\
$\cB(\Upsilon(2S) \to \mu^\pm \tau^\mp)$ & $(0.2 \pm 1.5 \pm 1.3) \times 10^{-6}$ ~;~~ $< 3.3 \times 10^{-6}$ (90\% C.L.) \cite{Lees:2010jk} \\
$\cB(\tau \to \mu \phi)$ & $< 8.4 \times 10^{-8}$ (90\% C.L.) \cite{Miyazaki:2011xe} \\
$\cB(J/\psi \to \mu^\pm \tau^\mp)$ & $< 2.0 \times 10^{-6}$ (90\% C.L.) \cite{Ablikim:2004nn} \\
 \hline
\end{tabular}
\end{center}
\caption{Measured values or constraints of the $2q2\ell$ observables
  that can significantly constrain the NP models.}
\label{tab:obs_meas}
\end{table*}
First performing a fit using only the minimal constraints we establish which models work and then in the second step we add LFV constraints to figure out the pattern of the NP couplings. 

The fit of S3 and U3 models to six minimal constraints gives $\chi^2=7.5$ and $10$ respectively. The degree of freedom(dof) in this case is two which is given by 6(No. of observables)- 4(No. of free parameters), implying a poor fit. The main reason for a bad fit
is found to be the upper bound on the $\mathcal{B}(B\to K \nu \bar \nu)$. 

\begin{figure}
\begin{center}
\includegraphics[width=0.4\textwidth]{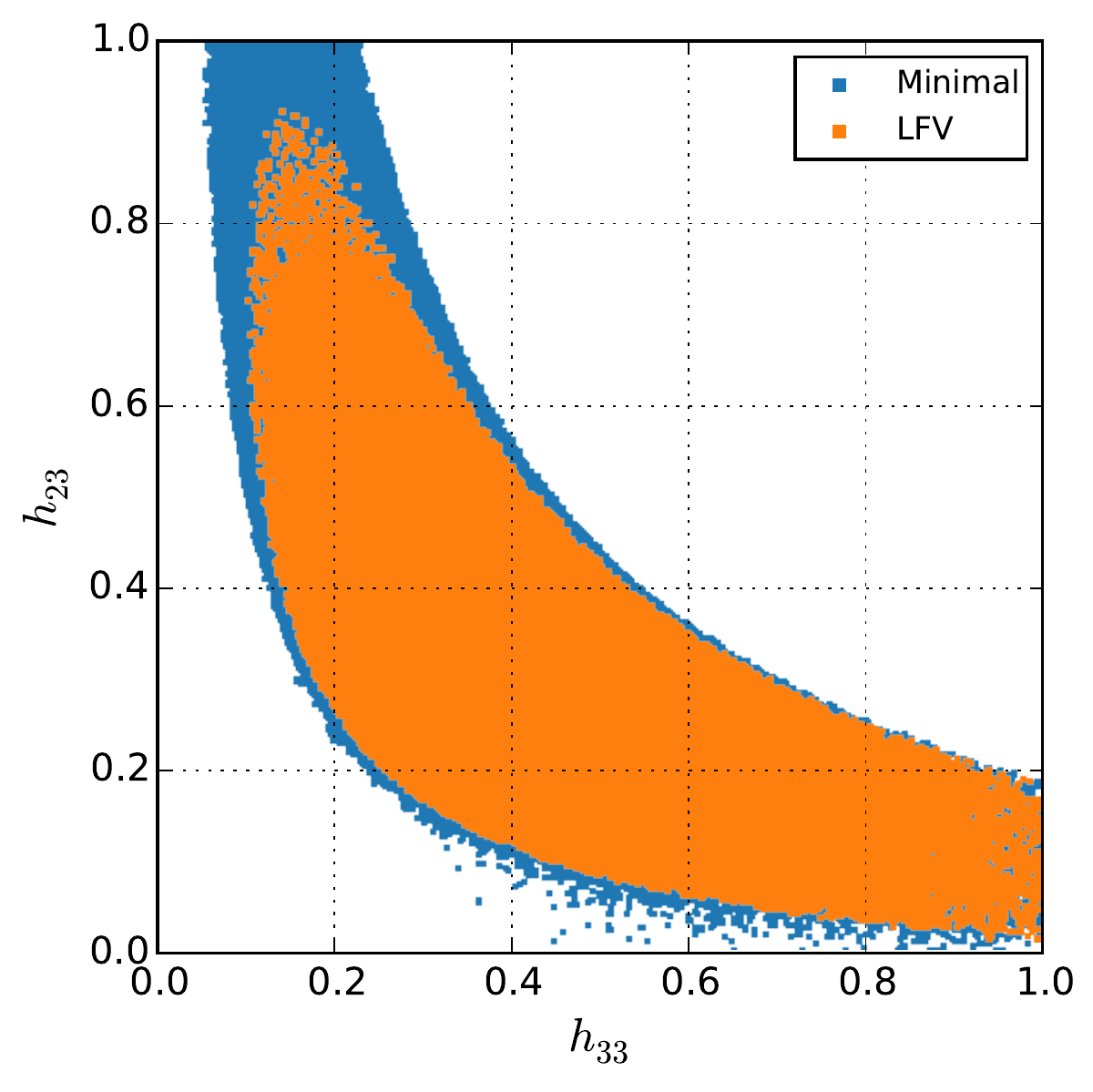}
\includegraphics[width=0.4\textwidth]{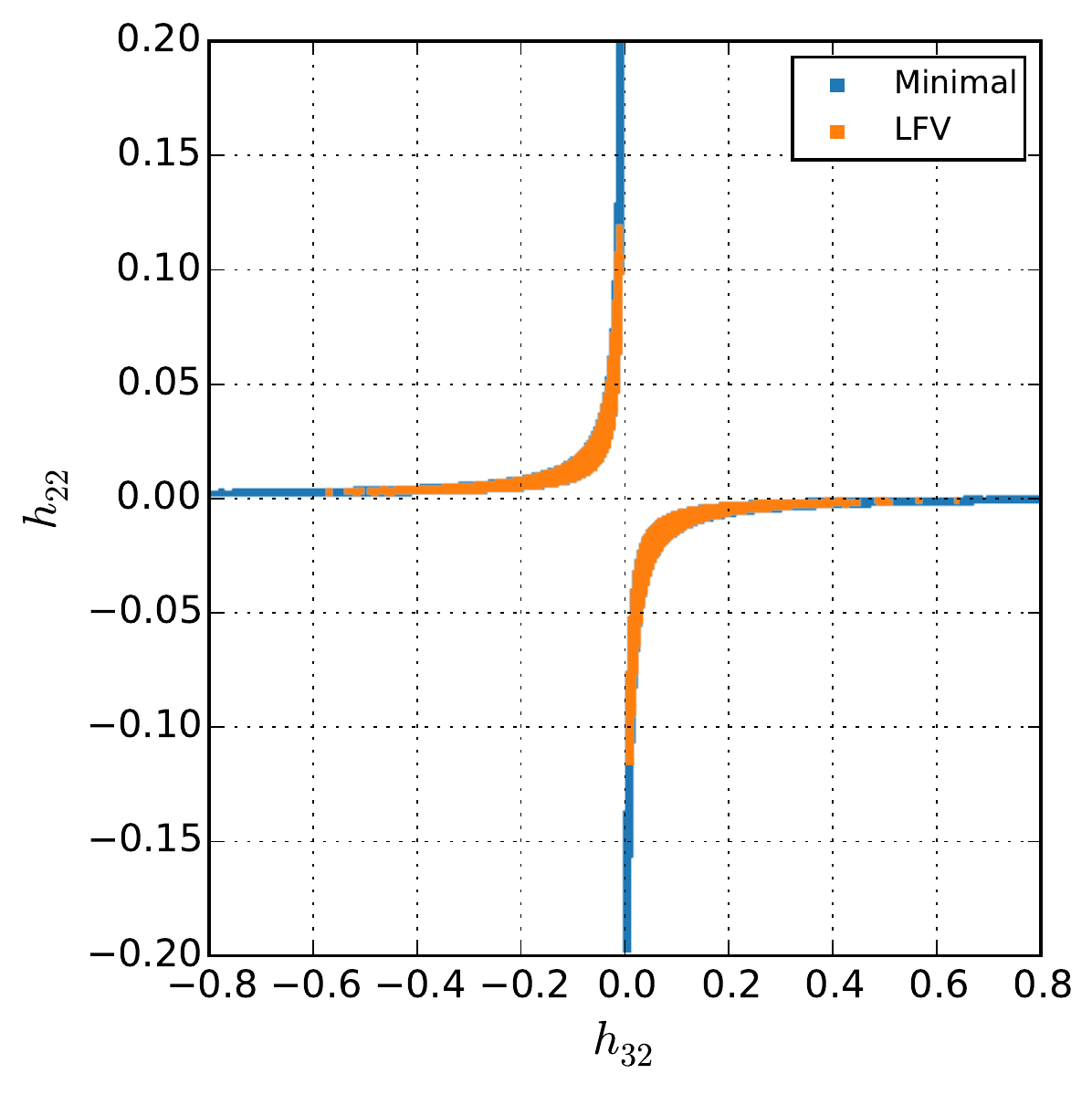}
\caption{Allowed 95\% C.L. regions in $h_{33}$-$h_{23}$ space (left
  plot) and $h_{32}$-$h_{22}$ space (right plot), for $M_{\rm LQ} = 1$
  TeV.  The regions are shown for a fit with only minimal constraints
  (blue) or minimal $+$ LFV constraints (orange).}
\label{fig:U1fit}
\end{center}
\end{figure}

Moving to U1 model, in this case the $b\to s\nu\bar \nu$ transitions are forbidden at the tree level. Including LFV observables we have total 9 constraints and 5 dof. In this case $\chi^2=5$ is found at the best fit point, which means a good fit. Hence, U1 model is able provide a combined explanation to the both kind of anomalies. 
But the minimal observables constrain only the product of the couplings $h_{32} h_{22}$ and $h_{33} h_{23}$. This is shown in the blue region of Figure \ref{fig:U1fit}.
Adding LFV constrains the allowed region as shown by the orange color. These constraints put the limits on the individual couplings as
\begin{equation}
|h_{22}| \le 0.12, ~|h_{32}| \le 0.7, ~|h_{23}\le 0.9|, ~|h_{33}|\ge 0.1.
\end{equation}
Furthermore, it was found that the LFV constraints prefer $h_{33}$ to be $\mathcal{O}(1)$  and a sizable $h_{23}\sim \mathcal{O}(0.1)$\cite{Kumar:2018kmr}. In the previous analyses( see e.g.\cite{CCO, Bordone:2017anc}) a large coupling to the third generation was introduced as a theoretical assumption which turn out to be a requirement by the LFV constraints here. 

Finally, the U1 model predicts
\begin{itemize}
\item the ratio $R_{\mathcal{B}(B\to \pi \ell \bar \nu)}^{\tau /\mu} \simeq R_{D^{(*)}} =1.2.$
\item for $R_{D^{(*)}} = 1.2$ the ratio $R_{ \mathcal{B}(B \to K\bar \nu \nu)}^{U1/SM} \simeq 1.3.$
\item for $R_{D^{(*)}} = 1.2$ the ratio $R_{\mathcal{B}(B\to K \tau^+ \tau^-)}^{U1/SM} \simeq 250 !$
\end{itemize}

\begin{figure}[htb]
\begin{center}
\includegraphics[width=0.4\textwidth]{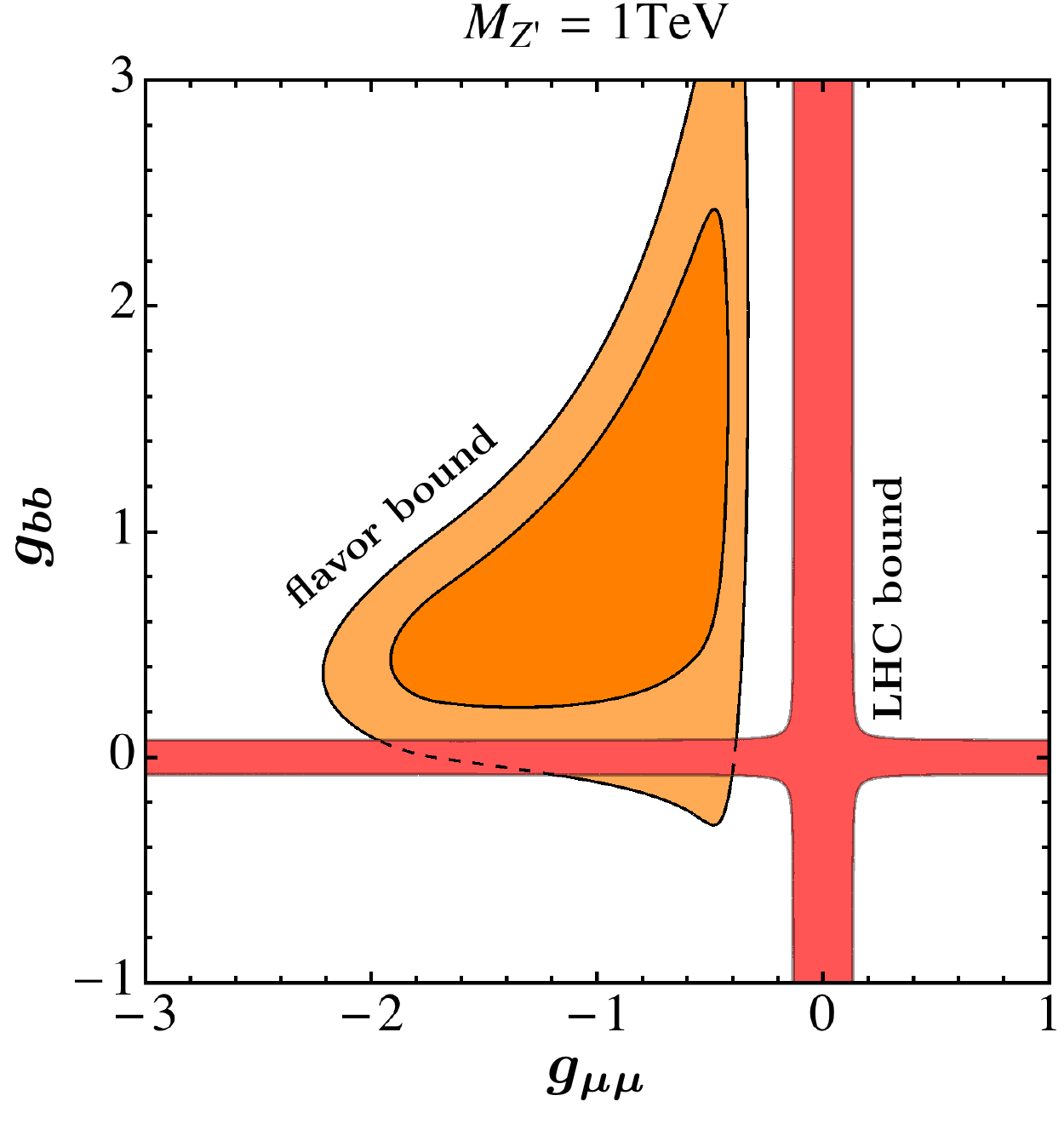}
\caption{Allowed regions in $(g_{bb},g_{\mu\mu})$ space from flavor
  and LHC constraints, assuming $M_{Z'} = 1$ TeV. The $1\sigma$ and
  $2\sigma$ flavor bounds are shown respectively in the dark and
  light orange regions. The 95\% C.L.\ LHC bound is shown in the red
  region.}
\label{LHCdimuon}
\end{center}
\end{figure}

Coming to VB model, it involves six couplings $(g_{\mu\mu}, g_{\mu \tau}, g_{\tau \tau})$ and $(g_{ss}, g_{sb}, g_{bb})$. An important difference from LQ models in this case is the presence of additional constraints due to $B_s$-mixing and purely leptonic decays such as $\tau \to 3\mu$ and $\tau \to \ell \nu \bar \nu$ at the tree level. It is found that $B_s$-mixing and $\mathcal{B}(\tau \to \ell \nu \bar \nu)$ constrain $g_{\tau \tau} \simeq \mathcal{O}(0.01-0.1)$. Because of this the NP effect on $b \to c \tau \bar \nu$ is very much limited and insufficient to accommodate $R_{D^{(*)}}$ anomalies. Therefore, the only option is to suppress the denominator of these observables which involve $b\to c \mu \bar \nu$. But then the direct searches of heavy vector boson at the LHC in the channel $b\bar b \to Z^\prime \to \mu^+ \mu^-$\cite{CMS:2016abv} turn out to be problematic. This is shown in Figure \ref{LHCdimuon}.

Therefore we conclude that:
\begin{itemize}
\item In case of leptoquarks, the S3 and U3 models are excluded by the constraint due to the upper bound on the $\mathcal{B}(B \to K \nu \bar \nu)$. The U1 model with a large coupling to the third generation and a sizable $h_{23}$ provides the combined explanation of the B-anomalies and predicts a large enhancement in the $\mathcal{B}(B\to K \tau^+ \tau^-)$. 
\item The VB model is excluded due to constraints coming from $B_s$-mixing, $\tau$-decays and direct searches in the dimuon channel at the LHC.
\end{itemize}

\begin{acknowledgments}
It was a wonderful experience to be at FPCP 2019, I thank the organizers of the FPCP 2019 for that . I also thank to David London and Ryoutaro Watanabe for collaborating in this work. This work was financially supported in part by NSERC of Canada.
\end{acknowledgments}

\bigskip 

\end{document}